\documentclass{iau}
\usepackage{graphicx}

\title[Roadmap on the theoretical work of BinaMIcS]%% give here short title %%
{Roadmap on the theoretical work of BinaMIcS}

\author[Mathis et al.]   %% give here short author list %%
{St\'ephane Mathis$^{1,2}$,
Coralie Neiner$^2$, 
Evelyne Alecian$^{2,3}$, 
Gregg Wade$^{4}$,
\and the BinaMIcS collaboration
}

\affiliation{
$^1$Laboratoire AIM Paris-Saclay, CEA/DSM-CNRS-Universit\'e Paris Diderot; IRFU
/SAp, Centre de Saclay, 91191 Gif-sur-Yvette Cedex, France\\ email: {\tt stephane.mathis@cea.fr}\\[\affilskip]
$^2$LESIA, UMR 8109 du CNRS, Observatoire de Paris, UPMC, Univ. Paris Diderot, 5 place Jules Janssen, 92195 Meudon Cedex, France\\[\affilskip]
$^3$UJF-Grenoble 1/CNRS-INSU, IPAG, UMR 5274, F-38041, Grenoble, France\\[\affilskip]
$^4$Dept. of Physics, Royal Military College of Canada, Kingston, K7K 7B4, Canada}

\pubyear{2013}
\volume{302}  %% insert here IAU Symposium No.
\pagerange{119--126}
\jname{Title of your IAU Symposium}
\editors{A.C. Editor, B.D. Editor \& C.E. Editor, eds.}
\begin{document}

\maketitle

\begin{abstract}
We review the different theoretical challenges concerning magnetism in interacting binary or multiple stars that will be studied in the BinaMIcS (Binarity and Magnetic Interactions in various classes of Stars) project during the corresponding spectropolarimetric Large Programs at CFHT and TBL. We describe how completely new and innovative topics will be studied with BinaMIcS such as the complex interactions between tidal flows and stellar magnetic fields, the MHD star-star interactions, and the role of stellar magnetism in stellar formation and vice versa. This will strongly modify our vision of the evolution of interacting binary and multiple stars.
\keywords{stars: magnetic fields, stars: binaries (including multiple): close}
%% add here a maximum of 10 keywords, to be taken form the file <Keywords.txt>
\end{abstract}

\firstsection % if your document starts with a section,
              % remove some space above using this command.
              
\section{The BinaMIcS project}

The BinaMIcS project has been awarded "Large Program" status with two high resolution spectropolarimeters: ESPaDOnS at the CFHT (Hawaii; PIs Alecian/Wade) for 604 hours over 4 years from 2013 to 2016, and Narval at TBL (France; PI Neiner) for 128 hours over 2 years from 2013 to 2014 (renewable for 2 more years). This large amount of time is being used to acquire an immense database of sensitive measurements of polarized and unpolarized spectra of spectroscopic binary (SB2) stars. The program includes 3 components: the detailed study of some known magnetic cool binaries, the detailed study of the few known magnetic hot close binaries, and a survey of a large number of hot binaries to search for magnetic fields. The detailed studies will allow us to obtain magnetic maps and test various models, while the survey part of the project will allow us to obtain statistical results. This database will be combined with new and archival complementary data (e.g. optical photometry, UV and X-ray spectroscopy) as well as theoretical studies, numerical simulations, and modeling (described hereafter), and applied to address the 4 main scientific objectives of the BinaMIcS project:
%\begin{itemize}
%\item 
i) what is the impact of magnetic fields during stellar formation, and vice versa;
%\item 
ii) how do tidally-induced internal flows impact fossil and dynamo fields;
%\item 
iii) how do magnetospheric star-star interactions modify stellar activity;
%\item 
iv) what is the magnetic impact on angular momentum exchanges and mass transfers.
%\end{itemize}

%\begin{figure}[t]
%\begin{center}
%\includegraphics[height=3cm]{collection_017_surface.ps} 
%\includegraphics[height=3cm]{collection_017_LSDIV.ps}\\
%\includegraphics[height=3cm]{collection_042_surface.ps} 
%\includegraphics[height=3cm]{collection_042_LSDIV.ps}\\
%\caption{Examples of the modelled surface (left), LSD Stokes I (middle) and V (right) profiles of $\beta$\,Cep fitted with pulsations and magnetic field, at two different phases.}
%\label{Neiner_fig1}
%\end{center}
%\end{figure}

\section{Roadmap on the theoretical work of BinaMIcS}

Magnetic fields are a crucial ingredient in a star's evolution, influencing its formation, the structure of its atmosphere and interior, as well as controlling the interaction with its environment. For binary stars, magnetism is even more significant, as magnetic fields in binary systems will be strongly affected by, and may also strongly affect, the transfer of energy, mass and angular momentum between the components. Therefore, the interplay between stellar magnetic fields and binarity has to be investigated in detail, from both the observational and theoretical point-of-view. First, the incidence and characteristics of magnetic fields are key parameters for understanding the physics of binaries. In higher-mass stars (above 1.5 $M_{\odot}$) the incidence of magnetic stars in binary systems provides a basic constraint on the detailed origin of the magnetic field, assumed to be a fossil remnant, and whether such strong magnetic fields suppress binary formation or are a result of mergers. Next, in low-mass stars, tidal interactions are expected to induce large-scale 3D shear and/or helical flows in stellar interiors that can significantly perturb the stellar dynamo. Similar flows may also influence the fossil magnetic fields of higher-mass stars. Finally, magnetically driven winds/outflows in cool and hot close binary systems have long been suspected to be responsible for their orbital evolution, while magnetospheric interactions have been proposed to enhance stellar activity. However, the crucial observational constraints required to test these hypotheses are, at present, nearly nonexistent, and dedicated theoretical studies are mandatory to bring the studies of binary (and multiple) stars to a new level of understanding. Within BinaMIcS, we will therefore study theoretically the complex interactions between tides and magnetic fields, i.e.:
i) how magnetic fields modify tidal flows described by \cite[Zahn 1977, Ogilvie \& Lin 2007, Le Bars et al. 2010, and Remus et al. 2012]{Zahn 1977,OgilvieLin2007,LeBarsetal2010,Remusetal2012} and the associated torques applied on each component;
ii) how flows driven by tides (as well as precession and libration) can modify dynamo mechanisms \cite[(e.g. Le Bars et al. 2011)], because of the angular momentum they transport ($\Omega$-effect) and of their helicity ($\alpha$-effect), and the stability of fossil fields;
iii) how such external mechanical forcings can compete with convective driving as a function of the mass ratio and of the separation of the components; iv) how do the combined action of tides and magnetic fields modify the orbital dynamics of binary stars and the evolution of their components \cite[(e.g. de Mink et al. 2013, Mathis \& Remus 2013, Song et al. 2013)]{deMinketal2013,MathisRemus2013,Songetal2013}. Moreover, we will study the complex MHD interactions between the components of binary stars, i.e.: i) the interactions between stellar wind emitted by each component and the related torques \cite[(e.g. Strugarek et al. 2012)]{Strugareketal2012}; ii) the magnetospheric interactions and the associated applied torques, helicity exchanges, and modifications of the magnetic activity of the components \cite[(e.g. Lanza 2012 and the contribution by S. Gregory in this proceeding)]{Lanza2012}. Results will also be applied to the study of star-planet interactions.
%and Cebron et al. 2012

\end{document}